\documentclass{article}

\begin{document}

\title{Prolegomena to any future Quantum Gravity}
\author{ John Stachel \\ Boston University}
\date{}
\maketitle

\begin{abstract}
I shall discuss some "conditions of possibility" of a quantum theory of gravity, stressing the need for solutions to some of fundamental problems confronting any attempt to apply some method of quantization to the field equations of  general relativity. 
\end{abstract}

\pagebreak

\tableofcontents

\pagebreak

\section{Introduction}

"Prolegomena" means "preliminary observations," and my title is meant to recall Kant's celebrated {\it Prolegomena to Any Future Metaphysics That Can be Claim to be a Science}. My words, like his:
\begin{quote}
are not supposed to serve as the exposition of an already-existing science, but to help in the invention of the science itself in the first place.
\end{quote}
To use another Kantian phrase, I shall discuss some "conditions of possibility" of a quantum theory of gravity, stressing the need for solutions to some of fundamental problems confronting any attempt to apply some method of quantization to the field equations of  general relativity (GR). Alternative approaches to quantum gravity (QG) are not discussed due to space-time (S-T) limitations (but see Stachel 2006). 

The first problem is the tension between "method of quantization" and "field equations of GR." The methods of quantization of pre-general-relativistic theories\footnote{In particular, non-relativistic  quantum mechanics (QM )based on Galilei-Newtonian S-T, special-relativistic quantum field theory based on Minkowski S-T, and quantum field theories in non-flat Riemannian S-Ts. But see Smolin 2005 for a discussion of topological QFT.} have been based on the existence of some fixed S-T structure(s); needed both for the development of the formalism and --equally importantly-- for its physical interpretation. This S-T structure provides a {\it fixed kinematical background} for any dynamical theory: The dynamical equations for particle or fields must be invariant under all automorphisms of the S-T symmetry group.

GR theory, on the other hand, is a {\it background-independent} theory, without any fixed, non-dynamical S-T structures. Its field equations are invariant under all differentiable automorphisms (diffeomorphisms) of the underlying manifold, which have no spatio-temporal significance until the dynamical fields are specified. In a background-independent theory, there is no kinematics independent of the dynamics\footnote{Ashtekar and Lewandowski 2002 note that "in interacting … [special-relativistic] quantum field theories, there is a delicate relation between quantum kinematics and dynamics: unless the representation of the basic operator algebra is chosen appropriately, typically, the Hamiltonian fails to be well-defined on the Hilbert space;" and go on to suggest that in GR one has the same "problem of choosing the 'correct' kinematical representation" (p. 51). By a "background independent kinematics" for GR they mean a "quantum kinematics for background-independent theories of connections." This phrase obscures the fact that, in a special-relativistic theory, the "basic operator algebra" is dictated by the symmetry group of the background S-T metric; while in GR it only emerges from the field equations after some canonical division into constraint and evolution operators.}.

GR and special relativistic quantum field theory (SRQFT) {\it do} share one fundamental feature that often is not sufficiently stressed: {\it the primacy of processes over states}\footnote{Baez 2006 uses cobordisms to emphasize this similarity.}. The four-dimensional approach, emphasizing processes in regions of S-T, is basic to both (see, e.g., Stachel 2006, Reisenberger and Rovelli 2002, DeWitt 2003). In non-relativistic quantum mechanics (QM), one can choose a temporal slice of S-T so thin that one can speak meaningfully of "instantaneous measurements" of the states of a system (see Micanek and Hartle 1996), but this is not so for measurements in SRQFT, let alone in GR (see, e.g. Bohr and Rosenfeld 1933, Bergmann and Smith 1982, Reisenberger and Rovelli 2002). 
	The breakup of a four-dimensional S-T region into lower-dimensional sub-regions  --in particular, into a one parameter family of three-dimensional {\it hypersurfaces}-- raises another problem. Such a breakup of processes in GR-- in particular, into the evolution of instantaneous states on a family of hypersurfaces-- is useful, perhaps sometimes indispensable, as a calculational tool. But no fundamental significance should be attached to such breakups, and results so obtained should be carefully examined for their significance from the four-dimensional, process standpoint (see, e.g., Nicolai and Peeters 2006). Since much of this paper is concerned with such breakups, it is important to emphasize this problem from the start.
	Perhaps the process viewpoint should be considered obvious in GR, but the use of three-plus-one breakups of ST in canonical approaches to QG (e.g., geometrodynamics and loop QG), and discussions of "the problem of time" based on such a breakup, suggest that it is not. The problem is more severe in the case of quantum theory, where the concepts of {\it state} and state function still dominate most treatments. But, as emphasized by Bohr and Feynman, the ultimate goal of any quantum-mechanical theory is the computation of the {\it probability amplitude} for some {\it process} leading from an initial preparation through some specified interaction(s) to a final registration (See Stachel 1997, 2005b, which include references to Bohr and Feynman).
	An important criterion, called "measurability analysis" (Bergmann and Smith 1982), based on "the relation between formalism and observation" (Reisenberger and Rovelli 2002), sheds light on this physical implications of any formalism. The possibility of the {\it definition} of some physically significant quantity within a theoretical framework should coincide with the possibility of its {\it measurement} in principle; i.e., by means of idealized measurement procedures consistent with that theoretical framework. Non-relativistic QM and in special relativistic quantum electrodynamics, have been tested by this criterion; and has been employed in testing proposals for what should be the fundamental quantities defined in QG (Bergmann and Smith 1982). This question is discussed in Section 4.
	The development of QM and SRQFT shows that the choice of variables in a classical theory and the mathematical methods use to describe processes undergone by these variables have play a major role in the determining the form, in which the transition to a quantized version of the classical theory can be effected, and sometimes even in determining the content of the resulting quantized theory\footnote{In the case of SRQFT, inequivalent representations of the basic operator algebra  are possible.}. Section 2 discusses three formalisms for Maxwell's theory and the corresponding quantizations, in particular the method based on Wilson loops that has led to the development of a background-independent quantization procedure (for a survey, see Smolin 2005). 
	Various possible choices of fundamental variables in GR are surveyed in Section 3, and various types of initial-value problems that may be posed for the evolution of these variables are discussed in Sections 5 and 6. Section 7 discusses classes of partially background-dependent GR S-Ts, having preferred Lie groups of symmetries, on which various "mini-" and "midi-superspace" toy models of QG are based. Quantization of the class of asymptotically flat S-Ts, in which a separation of kinematics and dynamics at null infinity allows application of conventional quantization techniques, is also discussed there. There is a brief Conclusion.

\section{Choice of Variables and Initial Value Problems in Classical Electromagnetic Theory}
In view of the analogies between electromagnetism (EM) and GR (see Section 3)-- the only two classical long-range fields transmitting interactions between their respective sources-- it is useful to begin by considering some of the issues that arise in QG in the simpler context of EM theory\footnote{This theory is simplest member of the class of gauge-invariant Yang-Mills theories, with gauge group $U(1)$; most of the following discussion could be modified to include the entire class (see, e.g., Carrion 2004).}.
There are two basic levels of analogy between the two. In the first, the electromagnetic four-vector potential is the analogue of the pseudo-metric tensor.
The second level of analogy between EM and GR is usually stressed in comparisons between Yang-Mills gauge fields and GR. Maxwell's theory is a $U(1)$ gauge theory, in which $A$ is the connection one-form, the analogue of the GR connection one-forms, and $F = dA$ is the curvature two-form, the analogue of the GR curvature two-forms(see Section 3 and 6, for the tetrad formulation of GR).
This first analogy has been pursued in two ways: 
\begin{enumerate}
\item The formulation of GR entirely in terms of the pseudo-metric tensor $g_{\mu\nu}$ (see Section 3) is analogous to the formulation of EM entirely in terms of potential four-vector $A_\mu$. In both, the field equations are of second order in the respective field variables. In the linearized approximation to GR, this analogy is formally very close: the linearized field equations for small perturbations $h_{\mu\nu}$ of the metric around the Minkowski metric $\eta_{\mu\nu}$ obey the same equations as does a special-relativistic, gauge-invariant massless spin-two field\footnote{For the important conceptual distinction between the two see Section 7, {\it Small Perturbations and the Return of Diffeomorphism Invariance}.}; while $A_\mu$ obeys the equations of a special-relativistic gauge-invariant spin-one field:
$$
(\eta^{\mu\nu}\partial^2_{\mu\nu})A_\kappa - \partial_\kappa (\partial_\mu\eta^{\mu\nu} A_\nu) = j_\kappa\, \,.
$$
These equations are invariant under the gauge transformation $A_\kappa \rightarrow A_\kappa + \partial_\kappa \chi$, where $\chi$ is a scalar field; and the divergence of the left-hand-side vanishes identically, so vanishing of the divergence of the right hand side (conservation of charge) is an integrability condition of the equations. These two features are related by Noether's second theorem (see Section 5).

\item The formulation of GR in terms of pseudo-metric and independently defined inertio-gravitational connection is analogous to the formulation of EM in terms of $A_\mu$, a one form $A$ (forms will be abbreviated by dropping indices)and an initially-independent field tensor $G_{[\mu\nu]}$, a two-form. The definition of the Christoffel symbols in terms of the metric tensor and its first derivatives is analogous to the definition of another EM field tensor $F_{[\mu\nu]}$, also a two form:

$$
F_{[\mu\nu]} = \partial_\nu A_\mu - \partial_\mu A_\nu \hspace{1cm} \left[ F = dA \right]  
$$

from which the first set of Maxwell equations follow:

$$
\partial_{[\kappa} F_{\mu\nu]} = 0 \hspace{1cm} \left[ dF=0 \right]
$$
The vacuum constitutive relations $[F = G]$ are analogous to the compatibility conditions equating the Christoffel symbols to the connection. The second set of Maxwell equations: 

$$
(G^{\mu\kappa} ) = j^\kappa \hspace{1cm} \left[ dG = j \right]
$$
where $j$ is the charge-current 3-form, are the analogue of the Einstein equation relating the affine curvature tensor to the stress-energy tensor. This analogy is especially close if GR is also expressed in terms of exterior differential forms (see Section 3). Writing the respective field equations in terms of exterior differential forms and putting  the theory into three-plus-one form, is the starting point of EM quantization in terms of Wilson loops, and of LQG (see, e.g., Smolin 2005). 
In EM, a family of parallel hyperplanes,  $t = const$ in some inertial frame of reference, is used. $A$  splits into the three-vector- and scalar-potentials, $\mathbf{A}$ and $\phi$. $F$ and $G$ split into $\mathbf{E}$ and $\mathbf{B}$ fields and $\mathbf{D}$ and $\mathbf{H}$ fields, respectively. In the rest frame of a linear, homogeneous isotropic medium\footnote{The rest frame of a material medium is a preferred inertial frame. In the case of the vacuum, a similar split may be performed with respect to any inertial frame.}, the constitutive relations become:

$$
\mathbf{D} = \epsilon \mathbf{E} \;\;\;\textbf{and} \;\;\; \mathbf{B} = - \mu \mathbf{H}
$$
$\epsilon$ and $\mu$ being the dielectric constant and permeability of the medium, and $\epsilon\mu = (n/c)^2$, where $n$ is its index of refraction and $c$ is the speed of light in the vacuum. The second order field equations split into one three-scalar and one three-vector equation:

$$
\frac{\partial}{\partial t}(div \mathbf{A})\,+\, (del)^2\phi = \rho, \;\;\;\;\;\; grad\, div\, \mathbf{A} - (del)^2 \mathbf{A} - \left(\frac{n}{c}\right)^2 \left( \frac{\partial^2\mathbf{A}}{\partial t^2}  \right) = j
$$

The first is a constraint equation. Using the gauge freedom to set $div \mathbf{A} = 0$ initially and $(del)^2\phi=\rho$ everywhere, the constraint insures that $(div\mathbf{A}) = 0$ holds everywhere, and the second equation, becomes the (vectorial) wave equation for $\mathbf{A}$.
\end{enumerate}

Going over from the second order Lagrangian to a first-order Hamiltonian formalism, one proceeds with canonical quantization, which may be carried out in either position- or momentum-space. The latter, based on the Fourier-transform of the potential and field components, leads to the Fock space representation of the quantized free field, useful for describing scattering experiments since the asymptotic in- and out-fields always may be treated as free. The momentum- and position-space representations are unitarily equivalent.
There is no "natural" breakup into three-plus-one in GR, so a spacelike hypersurface in the manifold is arbitrarily selected. Geometrodynamics takes the three-space metric of the hypersurface as "position" variables, and attempts canonical quantization; but it does not seem possible to give it a mathematically rigorous formulation (see Ashtekar and Lewandowski 2004). LQG takes a connection on the hypersurface as "position variables" (see Section 5); but a mathematically rigorous quantization seems to require the introduction of loop variables (see below).

Attempts to better understand LQG have inspired a similar approach to quantization of the EMF (see, e.g., Ashtekar and Rovelli 1992), based on loop integrals of field quantities around a closed curve on a hyperplane $t = const$. $\int_C \mathbf{A}$ is gauge-invariant and it follow from the definition of $\mathbf{E}$\footnote{If there are topological complications, the periods of $\int_C (grad\phi)$ must also be considered.} that $\int_C\mathbf{E} = d[\int_C\mathbf{A}]/dt$, so the former are taken as "position" variables, with the latter as the corresponding "velocities." This suggests the feasibility of a Feynman-type quantization of the theory, in which the classical S-T path of a such loop is an extremal of the timelike world tube bounded by the loops of $\int_C\mathbf{A}$ on the initial and final hyperplanes (see Baez 2006 for discussion of such cobordisms). The sum over all such paths would be used to calculate the quantum transition amplitude between the initial and final loops\footnote{See Reisenberger 1994 and Reisenberger and Rovelli 1997.}. 

More generally, loop integrals of A for {\it all possible types} of closed curves $C$ might be considered, leading to a Feynman-type quantization based on arbitrary spacelike initial and final hypersurfaces; and, by using loops lying on null hypersurfaces, null-hypersurface quantization techniques might be applicable.
But the standard approach to loop quantization is canonical (see, e.g., Ashtekar and Rovelli 1992): The momenta conjugate to $\int_C\mathbf{A}$ are $\int\int_S \mathbf{D}\cdot \mathbf{n} dS$, where $S$ is any two surface bounded by $C$\footnote{Stachel 1984 gives a Lagrangian density for arbitrary constitutive relations. When evaluated on t= const, the only term in the Lagrangian density containing a time derivative is $(\partial\mathbf{A}/\partial t)\cdot\mathbf{D}$, from which the expression for the momentum follows. If a non-linear constitutive relation is used, the difference between $\mathbf{D}$ and $\mathbf{E}$ becomes significant.}. The relation between  $\mathbf{D}$ (momentum) and $\mathbf{E}$ (velocity) is determined by the constitutive relations of the medium, the analogue of the mass relating a particle's momentum and velocity. In a four-dimensional formulation,.the "dual momenta" are the integrals $\int\int_S G$ over {\it any} two-surface $S$, suggest the possibility of extending the canonical loop approach to arbitrary spacelike and null hypersurfaces.
	While the position and momentum-space representations are unitarily equivalent; they are not unitarily equivalent to the loop representation. (see Carrión 2004). In order to secure unitary equivalence, it is necessary to introduce "smeared" loops (Varadarajan 2000, 2001)\footnote{The loops are "smeared" with a one parameter family of Gaussian functions over the three-space surrounding the loop.}. These results suggest that measurement analysis (see the Introduction) might show that "thickened" four-dimensional regions of S-T around a loop are needed for ideal measurement of loop variables. The implications of measurement analysis for loop quantization of GR also deserve careful investigation.

\section{Choice of fundamental variables in classical GR}
Two possible choices as the fundamental dynamical variables in GR are: the pseudo-metric and the linear connection, which are well known; and the conformal and projective structures, which have been much less explored (see Goenner 2005 Section 2.1, Geometries). The two are inter-related in a number of ways, only some of which will be discussed\footnote{Mathematically, these structures are best understood as G-structures of the first and second order; i.e., reductions of the linear frame bundle $GL(4, R)$ of the S-T manifold with respect to different subgroups. The metric structure and the volume structure are first order reductions of the frame bundle group with respect to the pseudo-orthogonal subgroup $SO(3,1)$ and unit-determinant subgroup $SL(4, R)$, respectively. The projective structure and the first order prolongation of the volume structure are  second order reductions of the frame bundle group. The interrelations between these structures follow from the relations between these subgroups  (see Sanchez-Rodriguez 2001).}.

\subsection{Metric and linear connection}
To this day, the coordinate components of the pseudo-metric\footnote{I shall often refer to it simply as "the metric" when there is no need to emphasize its Lorentzian signature.} field $g_{\mu\nu}$ are often taken as the only set of dynamical variables in GR, obeying second-order partial deifferential equations. Here, the metric tensor plays a dual role physically:
\begin{enumerate}
\item Through the invariant line element $ds$ between two neighboring points of the manifold, it determines the {\it chrono-geometry} of S-T ($ds^2 = g_{\mu\nu}dx_\mu dx_\nu$ for intervals that may be space-like, time-like or null). Since $ds$ is not a perfect differential, the proper time between two events depends on the time-like path between them. 
\item Its components also serve as the "potentials" for the Christoffel symbols, the Levi-Civita connection describing the {\it inertio-gravitational field}. It does so:
\begin{itemize}
\item directly, through the geodesic equation; describing the behavior of freely falling particles in the field. Metric geodesics are characterized as extremals of the interval: shortest for space-like, longest for time-like, or zero-length for null curves;
\item indirectly, through the role of the Riemann tensor $R_{[\kappa\lambda][\mu\nu]}$ in the equation of geodesic deviation, describing tidal gravitational forces.
\end{itemize}
\end{enumerate}
According to Einstein's equivalence principle, gravity and inertia are described by a single inertio-gravitational field and a reference frame can always be chosen locally ("free fall"), in which the components of the field vanish. Consequently, this field is represented by a symmetric linear connection $\Gamma^\kappa_{\mu\nu}$ in a four-dimensional formulation of Newtonian theory as well as in GR. 	
For this and other reasons, it is preferable to take both the pseudo-metric and connection as independent dynamical variables in GR. The connection still describes the inertio-gravitational field through the geodesic equation (affine geodesics now being the {\it straightest} paths in S-T).The affine curvature tensor $A^\kappa_{\lambda[\mu\nu]}$, plays a role in the affine equation of geodesic deviation analogous to that of the Riemann tensor in the corresponding metric equation. The field equations, which are now first order in the derivatives components of the metric and connection, can be derived from a Palatini-type variational principle; one set of equations fixes the compatibility conditions between metric and connection: the covariant derivative of the metric with respect to the connection must vanish,  ensuring that the connection is metric: straightest curves coincide with extremals; and  the Riemann tensor agrees with the affine curvature tensor when their components are raised or lowered with the metric tensor.
By introducing a tetrad of basis vectors $e_I$ and its dual co-basis of one-forms $e^I$, it is possible to write the components of the metric, connection and curvature tensor with respect to the tetrad in various ways. Recent progress in QG One has shown one way to be especially important. The co-basis of one forms is used to represent the chrono-geometry throught its Lorentz-invariant combination $g = \eta_{IJ} e^I e^J$ ($\eta_{IJ}$  is the Minkowski metric), and the affine connection and curvature tensor are represented by $SO(3,1)$-valued matrices of one-forms $\omega^I_J$ , and two-forms $R^I_J  = d\omega^I_J  + \omega^I_K  \wedge \omega^K_J$  , respectively (see, e.g., Rovelli 2004, or Stephani et al 2003). 
	This formulation enabled Ashtekar to put the field equations of GR into a form, in which they closely resemble those of Yang-Mills theory. Much recent progress in LQG is based on the introduction of the Ashtekar connection, a three-connection embodying all the information in the four-connection at each point of the hypersurface (see Section 6).

\subsection{Projective and Conformal Structures}
The metric and connection can each be further decomposed into two structures: The metric into the {\it conformal} (or {\it causal}) and the {\it four-volume-determining structures}; and the connection into the {\it projective} (or {\it path-determining}) and the {\it affine-parameter determining structures}. 
The conformal structure determines the {\it null wave fronts}, or dual {\it null rays}  of S-T. The projective structure determines the {\it preferred} ("{\it straightest}") {\it paths} of force-free monopole particles in S-T\footnote{An preferred affine {\it curve}, or geodesic, is parameterized by a preferred affine parameter; a preferred projective {\it path} is not so parameterized.}. Together, the two determine the metric of a given pseudo-Riemannian  S-T (Weyl 1921). Conversely, given conformal and projective structures on a manifold that obey certain compatibility conditions, the existence of a compatible metric is guaranteed (Ehlers, Pirani and Schild 1972). Compatibility between the two structures can be derived from a Palatini-type Lagrangian for GR taking both structures as dynamical variables. These structures might form the basis of an approach to QG that incorporates the insights of causal set theory (see Stachel 2006).

\section{The Problem of Quantum Gravity}
Measurability analysis (see the Introduction) of various possible dynamical variables in GR (see the previous section) may help in choosing a suitable maximal set of independent variables. Introduction of the quantum of action into such an analysis will limit joint measurability to compatible subsets, which could serve as a basis for quantization of GR. 
The formal representation of such measurement procedures involves the introduction of further, non-dynamical structures on the manifold, such as tetrads, bivector field, congruences of subspaces, etc, which are given a physical interpretation in the measurement context (see Rovelli 1991a, 1991b and Sections 5 and 6). This procedure is closely related to the question of possible choices of initial data and their evolution along congruences of subspaces (see Section 6). 
	It has been suggested that measurability analysis in GR be carried out at one or more of three levels: metric, connection and curvature (see the previous section):
	
{\it The pseudo-metric tensor}: Measurements of spatial or temporal integrals of various quantities along some curve $\int ds$ , or over spatial two-areas and three-volumes\footnote{This is especially important in view of the claim that quantized values of spatial two-areas and three-volumes are measurable (see, e.g., Ashtekar and Lewandowski 2004 or Rovelli 2004; for critical comments, see Nicolai and Peeters 2006). In particular the measurability of all two-surface integrals of the curvature two-forms, and not just over spatial two-surfaces, should be investigated.}, or spatio-temporal four-volume, could provide information about various aspects of the metric tensor. In a sense, all measurements ultimately reduce to the measurement of such entities\footnote{In the context of SRQFT, Kuhlmann 2006 notes: "[S]pace-time localizations can specify or encode all other physical properties."}. 
The Introduction and Section 2 present arguments suggesting that four-dimensional, process measurements are fundamental, and measurements of apparently lower-dimensional regions may actually be measurements of processes approximating such regions. Because of its fundamental importance, this question deserves further investigation. 
Measurement analysis might concern structures abstracted from the metric. For example, propagation of massless particles and fields involves only the conformal structure, while the four-volume of a process in some finite region in S-T involves only the four-volume structure.

{\it The affine connection}: The inertio-gravitational connection is not a tensor, but the choice of an appropriate physical frame of reference can serve to define an inertial connection, and the difference between two connections is a tensor. So a frame-dependent gravitational tensor can be defined and might be measurable, for example by measurement of the deviations of time-like geodesic curves  from purely inertial ones with respect to such a frame, together with proper time intervals along them. The mean value of quantum fluctuations around a classical connection also might be measurable.
Measurement analysis might concern structures abstracted from the affine structure. For example, if only measurements on the geodesic paths are made, the results depend only on the projective structure. 
Measurement analysis of "smeared" loop integrals of the connection one-forms over spatial and non-spatial S-T loops should be studied in connection with canonical and non-canonicals formulations of LQG.

{\it The Riemann or affine curvature tensor}: Measurement analysis of the components of the linearized Riemann tensor with respect to an inertial frame of reference has been studied in great detail (see DeWitt 1962, Bergmann and Smith 1982), and tentative lessons drawn concerning the full theory. Arguing that, in gauge theories, only gauge-invariant quantities should be subject to the commutation rules, DeWitt concluded that in GR measurement analysis should be carried out only at the level of the Riemann tensor (DeWitt 1962, 2003; see also Bergmann and Smith 1982). However, this conclusion neglects at least three important factors:

\begin{enumerate}
\item t follows from the compatibility of chrono-geometry and inertio-gravitational field in GR that measurements of the former can cast light on the latter. As noted above, the interval $ds$ between two neighboring events is already gauge invariant, as is its integral along any (non-null) world line. Indeed, all methods of measuring the Riemann tensor ultimately depend on the ability to measure such intervals, either a space-like $d\sigma$ or a time-like $d\tau$; both agree (up to a linear transformation) with the corresponding affine parameters on the geodesics.

\item The introduction of additional geometrical structures into the S-T manifold to model macroscopic preparation and registration devices introduces the possibility of additional gauge-invariant quantities (see Rovelli 1991a).

\item A geometric object may not be gauge-invariant, while some non-local integral of it is. As seen in Section 2, the electromagnetic four potential in electrodynamics is not gauge invariant, but its loop integrals are; indeed its non-vanishing periods form the basis for the Aharonov-Bohm effect. Similarly, at the connection level, the holonomies of the set of connection one-forms play an important role in the techniques used in LQG. (see, e.g., Ashtekar and Lewandowski 2004, Rovelli 2004). 
\end{enumerate}
In both EM and GR, one would like to have a method of loop quantization that does not depend on the singling out a family of spacelike hypersurfaces. For example, the various "problems of time" said to arise in the canonical quantization of GR seem to be more artifacts of the technique than genuine physical problems\footnote{That is, problems that arise from the attempt to attach some physical meaning to some {\it global time} coordinate introduced in the canonical formalism, the role of which in the formalism is purely as an ordering parameter with no physical significance (see Rovelli 1991b and Reisenberger and Rovelli 2002). The {\it real} problem of time is the role in QG of the {\it local or proper time}, which is a measurable quantity classically.}. The next section lists some non-canonical possibilities. 
Measurability analysis might involve some tensor abstracted from the Riemann tensor, such as the Weyl conformal curvature tensor. For example, measurability analysis of the Newman-Penrose formalism, based on the use of invariants constructed from the components of this tensor with respect to a null tetrad (see, e.g., Stephani et al 2003, Chapter 7), might suggest new candidates for QG dynamical variables. 

\section{The Nature of Initial Value Problems in General Relativity}
An initial value problem for a set of hyperbolic\footnote{Initial value problems are well-posed (i.e., have a unique solution that is stable under small perturbation of the initial data) for hyperbolic systems. It is the choice of Lorentz signature for the pseudo-metric tensor that makes the Einstein equations hyperbolic; or rather, because of their diffeomorphism invariance, only with the choice of an appropriate coordinate condition (e.g. harmonic coordinates) does the system of equations become hyperbolic.} partial differential equations on an n-dimensional manifold consists of two parts: specification of a set of {\it initial data} on some submanifold of dimension $d$ just sufficient to determine a unique solution; and construction of that solution, by showing how the field equations govern the {\it evolution} of the initial data along some $(n-d)$-dimensional congruence of subspaces. The problems can be classified in terms of the value of $d$ and nature of the initial submanifold, characteristic or non-characteristic; and the nature of the congruence of $(n-d)$-dimensional congruence of subspaces.
In four-dimensional S-T, there are only two possibilities:
\begin{itemize}
\item $d= 3$: Initial hypersurface(s) and evolution along a vector field (three-plus-one problems). 
\item $d= 2$: Two dimensional initial submanifolds, and evolution along a congruence of two-dimensional subspaces (two-plus-two problems). 
\end{itemize}
\subsection{Constraints Due to Invariance Under Function Groups}
If a system of $m$ partial differential equations for m functions is derived from a Lagrangian invariant up to a divergence under some transformation group depending on $q$ functions of the $q$ independent variables ($q\leq m$), then by Noether's second theorem (see, e.g., Winitzki 2006) there will be $q$ identities between the $m$ equations. Some of the $m$ functions are redundant when initial data is specified on a (non-characteristic) hypersurface, and the set of $m$ field equations splits into $q$ {\it constraint equations}, which need only be satisfied initially, and $(m-q)$ evolution equations. If the latter are satisfied off the initial hypersurface, the former will also be as a consequence of the identities. 
	The ten homogeneous ("empty space") Einstein equations for the ten components of the pseudo-metric field as functions of four coordinates are invariant under the four-parameter diffeomorphism group; hence, giving rise to the four contracted Bianchi identities between them. In the three-plus-one initial value or Cauchy problem on a spacelike hypersurface (see Friedrich and Rendell 2000), the ten field equations split into four constraints and six evolution equations. The ten components of the pseudo-metric are a highly-redundant description of the field, which has only two degrees of freedom per S-T point. 
	Isolation of these "true" degrees of freedom" of the field is a highly non-trivial problem. Quantization of the theory could take place either after their isolation and elimination of the constraints, which (apart from some simple toy models-- see Section 7) has not been achieved; or before their isolation, in which case superfluous degrees of freedom are first quantized and then eliminated via the quantized constraints, as in loop quantum gravity  (see Ashtekar and Lewandowski 2004, p. 51).
	
Initial value problems may serve to determine various ways of defining complete (but non- redundant) sets of dynamical variables. Each problem determines the sort of non-dynamical structures needed to make possible the definition of such a set and suggests appropriate measurement procedures for them (see Section 4). Such classical results also provide important clues about choices of variables for possible transition to QG. 
For canonical quantization, these questions have been extensively studied. The analogy between the probability of an ensemble of classical trajectories and the corresponding Feynman probability amplitude (see, e.g., Stachel 2005b) suggests the feasibility of a direct Feynman-type formulation of QG (see, e.g., DeWitt 2003, Preface). One can use initial value formulations as a method of defining ensembles of trajectories, based on specification of half the maximal classical initial data-- and then the maximal quantum-theoretically compatible data-- on an initial and final hypersurface. In Section 2, this possibility was discussed for the loop formulation of electromagnetic theory. 	

In GR, such an approach has been investigated for connection formulations of the theory, in particular for the Ashtekar loop variables in Reisenberger and Rovelli 1997, who state: "Spin foam models are the path-integral counterparts to loop-quantized canonical theories."\footnote{See Baez 2006 for the analogy between spin foams in GR and processes in quantum theory based on cobordisms.}

These methods of investigating the transition from classical to quantum theory are based on Cauchy initial value problems. Whether analogous methods can be based on two-plus-two initial value problems is an open question. Another possible starting point for canonical quantization is the {\it null-hypersurface} initial value problem. {\it Characteristic hypersurfaces} of a set of hyperbolic partial differential equations are those, on which no amount of initial data suffices to determine a unique solution; and in GR, null hypersurfaces are characteristics. 

But data can be specified on a pair of intersecting null hypersurfaces fixing a unique solution in the S-T region to the future of both (see, e.g., d'Inverno and Stachel 1978, Winicour 2005). There is a sort of two-for-one tradeoff between the initial data needed on a {\it single} Cauchy hypersurface and a {\it pair} of null hypersurfaces. While "position" and "velocity" variables must be given on the Cauchy hypersurface, only "position" variables need be given on the pairof null hypersurfaces. 

Various approaches to null hypersurface quantization have been applied. One of the two null hypersurfaces can be chosen as future or past null infinity $\Im^{\pm}$ (for a discussion of $\Im$ as boundary of an asymptotically flat S-T and quantization on it, see Section 7), and one method of quantization combines an asymptotic and another null hypersurface (Frittelli et al 1997). 

\subsection{Non-Dynamical Structures and Differential Concomitants}
GR is diffeomorphism-invariant, interpreted as invariance under the group of active diffeomorphisms of the points of the underlying manifold (it is trivial that all results are independent of a {\it passive} change of [local] coordinate system). Any additional non-dynamical structures needed to formulate initial value problems for the dynamical variables of the theory should be introduced by
geometrical, coordinate-independent, definitions. In particular, evolution of the dynamical variables should not involve the introduction of a preferred "time" coordinate\footnote{Subsequent introduction of a coordinate system adapted to some geometrical structure is often useful for calculations. But coordinate-dependent descriptions of an initial value problem, by doing tacitly what should be done explicitly, often create confusion.}. 
The dynamical fields include the pseudo-metric and inertio-gravitational connection, or structures abstracted from them (see Section 3), so any differential operators introduced to describe their evolution should be independent of any metric or connection\footnote{Similarly, if the conformal and projective structures are taken as the primary dynamical variables, the operators should be independent of these structures.}. That is, the operators must be {\it differential concomitants} of the dynamical variables and any non-dynamical structures introduced\footnote{A differential concomitant of a set of geometric objects is a geometric object formed from algebraic combinations of the set and their partial derivatives.}; the most commonly used ones are the {\it Lie derivative} $\mathcal{L}_v\Phi$ of a geometric object $\Phi$ with respect to a vector field $v$, and the exterior derivative $d\omega$ of a form $\omega$\footnote{Or, equivalently, the "curl" of a totally antisymmetric covariant tensor and "divergence" of its dual contravariant tensor density.} (see, e.g., Stephani et al, 2003, Chapter 2). Various combinations andgeneralizations of both are known, such as the Schouten-Nijenhuis and Frölicher-Nijenhuis brackets, and have been --or could be-- used in the formulation of various initial value problems in GR.

\section{Congruences of Subspaces and Initial-Value Problems in GR}
Initial value problems in GR involve:

\begin{enumerate}

\item choice of an initial submanifold and a complementary congruence of subspaces\footnote{"Complementary" meaning that the total tangent space at any point can be decomposed into the sum of the tangent spaces of the initial submanifold and of the complementary subspace.}; 

\item choice of a differential concomitant to describe the evolution of the initial submanifold and of the dynamical variables (see next point) via the congruence of complementary subspaces;

\item choice of a set of dynamical variables, usually related to the pseudo-metric and the affine connection, and their splitting by projection onto the initial submanifold and the complementary subspace;

\item breakup of the field equations into constraint equations on the initial submanifold and evolution equations along a congruence of complementary subspaces.

\end{enumerate}
	
The non-dynamical steps 1) and 2) will be discussed in this subsection, the dynamical ones 3) and 4) in the next.	

As discussed above, in GR there are only two basic choices for step 1): three-plus-one or two-plus-two splits\footnote{Various sub-cases of each arise from possible further breakups, and I shall mention a few of them below.}. But two further questions arise: is the initial submanifold {\it null}, and is the congruence of subspaces {\it holonomic}?	

In the three-plus-one case, a sufficiently smooth vector field is always holonomic (curve-forming); but in the two-plus-two case, the tangent spaces at each point of the congruence of two-dimensional subspaces may not fit together holonomically to form submanifolds. 	

In any theory involving a pseudo-metric (or just a conformal structure), either the initial submanifold or the complementary subspace may be {\it null}, i.e., tangent to the null cone\footnote{}. In a null tangent space of dimension $p$ there is always a unique null direction, so the space splits naturally into $(p-1)$- and $1$-dimensional  subspaces. The $(p-1)$-dimensional subspace is not-unique but is always spacelike. 	
	
A non-null tangent space of dimension $p$ in a manifold of dimension $n$ with pseudo-metric has a unique {\it orthogonal} tangent space of dimension $(n-p)$; so there are orthogonal projection operators onto	the $p$- and $(n-p)$-dimensional subspaces. The evolution of initial data on a spacelike p-dimensional submanifold is most simply described along a set of $(n-p)$-orthonormal vectors spanning the orthogonal congruence of subspaces (or some invariant combination of them -- see the next subsection). Otherwise, lapse and shift functions must be introduced, relating the congruence of subspaces used to the orthonormal congruence.
	
By definition, null vectors are orthogonal to themselves, so the construction of an orthonormal subspace fails for null surface-elements. On a null hypersurface, there is no orthonormal, so the null-initial value problem is rather different (see the previous and the next subsections). A similar analysis of two-plus-two null versus non-null initial value problems has not made, but one would expect somewhat similar results.

\subsection{Vector Fields and Three-Plus-One Initial Value Problems}
In the Cauchy problem, the use of a unit vector field $n$ normal to the initial hypersurface leads to the simplest formulation. $\mathcal{L}_n\Phi$ the Lie derivatives w.r.t. this field applied to the chosen dynamical variables $\Phi$ provides a natural choice of differential concomitants to define their "velocities" in the Lagrangian and their "momenta" in the Hamiltonian formulation of the initial-value problem; and higher order Lie derivatives can be used to treat their evolution in the unit normal direction. If $\mathcal{L}_v$ with respect to another vector field $v$ is used, the relation between $v$ and $n$ must be specified in terms of the lapse scalar $\rho$ and the shift vector $\sigma$,

$$
v = \rho\,n\,+\,\sigma
$$
The Cauchy problem for the Einstein equations has a major drawback: The initial data on a space-like hypersurface are subject to four constraint equations (see Section 5), which must be solved in order to find a pair of "true observables," freely specifiable as "positions" and "velocities" initially, the evolution of which off the initial hypersurface should be uniquely determined by a pair of coupled, nonlinear field equations. Only in certain highly idealized cases, such as cylindrical waves (see Section 7) can this program be carried out with locally-defined variables. In general, quantities expressing the degrees of freedom and the equations governing their evolution are highly non-local and can only be specified implicity, for example in terms of the conformal two structure coordinate and velocity (see d'Inverno and Stachel 1978). In this respect, things are somewhat better for the null and two-plus-two initial value problems.

A null hypersurface is naturally fibrated by a null vector field, and the initial data can be freely specified in a rather "natural" way on the family of transvecting space-like two surfaces. The projection of the metric tensor onto a null hypersurface is a degenerate three-metric of rank two, providing a natural metric for the two-surfaces. The halving of initial data (also discussed above)means that only two quantitities per point of the initial null hypersurface (the "positions") can be specified, leading to considerable simplification of the constraint problem; the price paid is the need to specify initial data on two intersecting null hypersurfaces. One way to do this is to start with a spacelike two-surface and drag it along {\it two} independent congruences of null directions, resulting in two families of spacelike two-surfaces on two null hypersurfaces. The initial data can be specified on both families of two-surfaces, generating a {\it double-null} initial value problem is generated. But the same data could also be specified on the initial two-surface, together with all of its Lie derivatives with respect to the two congruences of null vectors  This approach provides a natural transition to two-plus-two initial value problems.

\subsection{Simple Bivector Fields and Two-Plus-Two Initial Value Problems}
In the two-plus-two case, one starts from a space-like two-manifold, on which appropriate initial data may be specified freely (see d'Inverno and Stachel 1978); the evolution of the data takes places along congruence of time-like two surfaces that is either orthonormal to the initial submanifold, or is related to the orthonormal subspace element by generalizations of the lapse and shift functions (see Rosen 1987). The congruence is holonomic, and a pair of commuting vector fields\footnote{They are chosen to commute, so that all results are independent of the order, in which draggings along one or the other vector field takes place.} spanning may be chosen, and evolution off the initial two-manifold carried out using Lie derivatives w.r.t. the two vector fields They may be chosen either as one time-like and one space-like vector, which leads to results closely related to those of Cauchy problem\footnote{If one drags the spacelike-two surface first with the spacelike vector field, one gets an initial spacelike hypersurface}; or more naturally as two null vectors, which, as noted above, leads to results closely related to the double-null initial value problem. It is also possible to avoid such a breakup of the two-surfaces by defining a differential concomitant depending on the metric of the two-surface elements (see Stachel 1987).

\subsection{Dynamical Decomposition of Metric and Connection}
A $p$-dimensional submanifold in an $n$-dimensional manifold can be "rigged" at each point with a complementary $(n-p)$-dimensional subspace "normal" to it\footnote{The word "normal" here is used without any metrical connotation. "Transvecting" would be a better word, but I follow Weyl 1922.}. Every co- or contravariant vector at a point of the surface can be uniquely decomposed into tangential and normal components; and hence any tensor can be similarly decomposed.	

{\it Metric}: If the concept of "normal subspace" is identified with "orthogonal subspace\footnote{This identification excludes the case of null submanifolds.}," the metric splits into just two orthogonal components\footnote{Here, and as far as possible below, I shall avoid the use of indices where their absence is not confusing.}: 

$$
g\,=\,'g\,+''g \hspace{1cm} 'g \cdot ''g \,=\,0
$$
where $'g$ refers to the $p$-dimensional submanifold, and $''g$ refers to the $(n-p)$-dimensional orthogonal rigging subspace. The properties of these subspaces, including whether they fit together holonomically to form submanifolds, can all be expressed in terms of $'g$ , $''g$ and their covariant derivatives (see Stachel 1987); and all non-null initial value problems can be formulated in terms of such a decomposition of the metric. It is most convenient to express 'g in covariant form, to extract the two dynamical variables from it, and express "g in contravariant form, and to use to the form the differential concomitant needed to describe the evolution of the dynamical variables. "g is the pseudo-rotationally invariant combination of any set of pseudo-orthonormal basis vectors spanning the subspace, and one may form a similarly invariant combination of their Lie derivatives\footnote{The simple multivector formed by taking the antisymmetric exterior product of the basis vectors is also invariant under a pseudo-rotation of the basis, and a similarly invariant exterior product of their Lie derivatives may also be used.}.
In view of the importance of the analysis of the affine connection and curvature tensors in terms of one- and two-forms, respectively, it is important in carrying out the analysis at the metric level, to include representations based on tetrad vector fields and the dual co-vector bases, spanning the $p$-dimensional initial surface and the $(n-p)$-dimensional rigging space by corresponding numbers of basis vectors.

{\it Connection}: An n-dimensional affine connection can be similarly decomposed into four parts with respect to a $p$-dimensional submanifold and  complementary "normal" $(p-n)$-dimensional subspace (see note 20 and Weyl 1922).Consider an infinitesimal parallel displacement using the $n$-connection in a direction tangential to the submanifold. The four parts are: 

\begin{enumerate}
\item {\it The surface affined or $(t, t)$ connection}: The $p$-connection on the submanifold that takes a tangential vector into the tangential component of the parallel-displaced vector.
\item {\it The longitudinal or $(t, n)$ curvature}\footnote{The use of "curvature" here is a remindher of its meaning in the Frenet-Serret formulas for a curve, and has nothing to do with the Riemannian or affine curvature tensors.}: The mapping taking a {\it tangential} vector into the infinitesimal {\it normal} component of its parallel-displaced vector.
\item {\it The $(n, n)$ torsion}\footnote{Note this use of "torsion" has nothing to do with an asymmetry in the connection. All connections consaiderd in this paper are symmetric.}: The linear mapping taking a {\it normal} vector into the {\it normal} component of its \item parallel-displaced vector.
\item {\it The transverse or $(n, t)$ curvature}: The linear mapping that taking a {\it normal} vector into the infinitesimal {\it tangential} component of its parallel-displaced vector.
\end{enumerate}
Using covectors, one gets a similar decomposition of the matrix of connection one-forms.	
Such decompositions of metric and connection can be used to investigate $(3 + 1)$ and $(2 + 2)$ decompositions of the first order form of the field equations and the compatibility conditions between metric and  affine connection, and in first order formulations of initial value problems.
	If the $n$-connection is {\it metric}, then "normal" now has the additional meaning of "orthogonal" (see above). The $(t, t)$ surface affine connection is now the one compatible with the surface metric; the $(t, n)$ $(n, t)$ curvatures are equivalent; and the $(n, n)$ torsion reduces to an infinitesimal rotation.
	On a {\it hypersurface}, the torsion vanishes, and the $(t, n)$ and $(n, t)$ curvatures are both equivalent to the {\it second fundamental form} of the hypersurface. The Ashtekar connection combines the $(t, t)$ and $(n,t)$ curvatures into a single connection. 
	Extension of the Ashtekar variables, or some generalization of them, to null hypersurfaces is currently under investigation\footnote{For a review of some results of a generalization based on null hypersurfaces, see Robinson 2003.  For null Ashtekar variables, see D'Inverno et al 2006.}.
	In the two-plus-two decomposition, the $(n, n)$ rotation is present and there are a pair of second fundamental forms. For a formulation of the two-plus-two initial value problem when the metric and connection are treated as independent before imposition of the field equations, see Rosen 1987. Whether some analogue of the Ashtekar variables can be usefully introduced in this case remains to be studied.
	
\section{Background Space-Time Symmetry Groups}	
The isometries of a four-dimensional pseudo-Riemannian manifold are	characterized by the dimension $m\leq 10$ of its isometry group (or group of automorphisms 
or motions) and the dimension $o\leq  min (4, m)$ of this group's highest-dimensional orbits 
(see, e.g., Stephani et al 2003, Hall 2004). There are two extreme cases:

	The maximal symmetry group $(m =10, o = 4)$. Minkowski S-T is the unique Ricci-flat S-T in this group. Its isometry group is the Poincaré or inhomogeneous Lorentz group, which acts transitively on the entire S-T manifold. Special-relativistic field theories involve field equations that are invariant under this symmetry group. They are the most important example of {\it background-dependent} theories (see Introduction). 
	
	At the other extreme is the class of generic metrics, i.e., S-Ts with no non-trivial symmetries $(m =0, o = 0)$. Field equations that are required to be invariant under the group of all diffeomorphisms of the underlying four-dimensional differentiable manifold will include a class of solutions that is a subclass of this class\footnote{This global, {\it active} diffeomorphism group must not be confused with the groupoid of {\it passive}, local coordinate transformations. Nor must the trivial freedom to carry out active diffeomorphisms acting on {\it all structures} on the manifold, including whatever fixed background metric field (such as the Minkowski metric) may be present, be confused with the existence of a {\it subgroup} of such diffeomeorphisms that constitutes the symmetry group of this background metric. Only the latter has any physical significance and is discussed in this section. }. Such theories, not  involving any background S-T structures, are called {\it background-independent} theories. GR is such a theory. 

\subsection{Non-Maximal Symmetry Groups and Partially-Fixed Backgrounds}	
Between these two extremes lie all solutions to a background-independent theory that are required to preserve some fixed {\it non-maximal symmetry group}. We shall speak of each class of solutions with such a given symmetry group as being a {\it partially-fixed} background theory. 
	Actually no generic solution to the exact Einstein equations is known. Only the a priori imposition of such a partially fixed background symmetry group enables construction of explicit solutions --or rather classes of solutions-- to the Einstein equations (see, e.g., Stephani et al 2003).
In each such case, the background-dependent symmetry group determines a portion of the pseudo-metric tensor field non-dynamically. Only the remaining portion is unrestricted, and hence obeys a reduced set of dynamical Einstein field equations. One must examine each case to see how much freedom remains in the reduced  class of solutions to these equations.
Considerable work has been done on two classes of solutions:
\begin{enumerate}	
\item	The "mini-superspace" cosmological solutions, in which so much symmetry is imposed that only functions of one evolution parameter (the "time") are subject to dynamical equations (see, e.g., Ashtekar, and Uggla 1993a, 1993b). Quantization here resembles the quantization of a dynamical system of particles rather than waves, and does not seem likely to shed much light on the generic case.
\item The "midi-superspace" (see Torre 1998) solutions, notably the cylindrical wave metrics (see Bicák 2000), for which sufficient freedom remains to include both degrees of freedom of the gravitational field. In an appropriately adapted coordinate system, they can be represented by a pair of fields obeying non-linear, coupled scalar wave equations in two-dimensional flat S-T. In addition to static and stationary fields, the solutions include gravitational radiation fields having both states of polarization. Their quantization can be carried out as if they were two-dimensional fields. But, of course, the invariance of any results under the remaining diffeomorphism freedom must be carefully examined, as well as possible implications for the generic case.	
\end{enumerate}
	Stephani et al (2003) discuss solutions to the Einstein equations having groups of motions with null and non-null orbits, so it should be possible to study the quantization of such metrics in a systematic way.
	
\subsection{Small Perturbations and the Return of Diffeomorphism Invariance}	
	
While the fiber space of all four-metrics over a manifold is itself a manifold, the space of all four-geometries is {\it not} a manifold\footnote{}. It is a {\it stratified manifold}, partitioned into {\it slices}; each of which consists of all geometries with the same symmetry group, starting with the manifold of generic geometries having no non-trivial symmetries, and ending with the metrics having the maximal symmetry group.

But, if a metric is perturbed, unless the perturbations are restricted to lie within some symmetry group, the smallest perturbation of a geometry with non-trivial symmetry group takes a geometry into the generic slice of the stratified manifold. This observation is often neglected; in particular, when perturbation-theoretic quantization techniques developed for special relativistic theories are applied to the perturbations of the Minkowski solution. Diffeomorphisms of such a perturbation cannot be treated as pure gauge transformations on a fixed background Minkowki S-T, but modify the entire causal and inertio-gravitational structure of S-T (see, e.g., Doughty 1990, Chapter 21). This seems to be the fundamental reason behind the problems in applying special relativistic quantization techniques to such perturbations.	
	
\subsection{Asymptotic symmetries}	
An important class of solutions to the field equations lacks global symmetries, but has asymptotic symmetries as infinity is approached in null directions, allowing for their asymptotic quantization (see Komar 1973, Section VI, and Ashtekar 1987). Imposition of certain conditions on the behavior of the Weyl tensor in the future or past null limit allows conformal compactification of a large class of S-Ts (Penrose 1963) by adjoining boundary null hypersurfaces, $\Im^{\pm}$ (read "scri plus and scri-minus"), to the S-T manifold. Both $\Im^{\pm}$ have a symmetry group that is {\it independent} of particular dynamical solutions to the field equations in this class. Thus, there is a separation of kinematics and	dynamics on $\Im^{\pm}$, and a quantization based on this asymptotic symmetry group can be carried out in more or less conventional fashion. 
	"More or less" because the asymptotic symmetry group, the Bondi-Metzner-Sachs (BMS) group, is not a finite-parameter Lie group  like the Poincaré group, but includes four-called "supertranslation," functions that depend on two "angular" variables. Nevertheless, asymptotic gravitons may be defined as representations of the BMS group, independently of how strong the gravitational field is in the interior of the S-T (Ashtekar 1987). 

\section{Conclusion}
This paper has discussed only some possible approaches to quantization of the field equations of GR. In this sense, and in spite of its emphasis on background-independent techniques, the paper is rather conservative, ignoring such promising avenues of research as causal set theory, causal dynamic triangulation, twistor theory; and attempts to derive S-T structures from radically different entities, such as the symmetries of coherent states in quantum information theory
	It is by no means certain that any conservative approach will lead to the most fruitful fusion of quantum theory and GR --indeed, it is even probable that it will not. But until some approach has been developed leading to a consensus in the QG community, every conservative approach deserves to be explored to its limits, if only to draw lessons for a better alternative approach from the limited successes and ultimate failure of such attempts.

\section*{Acknowledgements} I thank Dr. Mihaela Iftime for reading an earlier draft of this paper and making several suggestions for improvement of this version.

\end{document}